\documentclass{emulateapj}

\newcommand{\snug}{$0.351 ^{+0.210}_{-0.139} \pm 0.020 $}
\newcommand{\snur}{$0.288 ^{+0.172}_{-0.114} \pm 0.018 $}
\newcommand{\snui}{$0.229 ^{+0.137}_{-0.091} \pm 0.014 $}
\newcommand{\snuz}{$0.186 ^{+0.111}_{-0.074} \pm 0.010 $}
\newcommand{\snuM}{$0.098 ^{+0.059}_{-0.039} \pm 0.009 $}
\newcommand{\snuB}{$0.36 ^{+0.22}_{-0.14} \pm 0.02 $}

\shorttitle{SN~Ia Rate in Low-Redshift Galaxy Clusters}
\shortauthors{Sharon et al.}

\begin{document}

\title{Supernovae in Low-Redshift Galaxy Clusters: \\
the Type-Ia Supernova Rate}

\author{Keren Sharon}
\affil{School of Physics and Astronomy, Tel Aviv University, Tel Aviv 69978,
Israel.} 

\author{Avishay Gal-Yam\footnote{Hubble Fellow}}
\affil{California Institute of Technology, Mail Code 105-24, Pasadena, CA
91125.} 

\author{Dan Maoz}
\affil{School of Physics and Astronomy, Tel Aviv University, Tel Aviv 69978,
Israel.} 

\author{Alexei V. Filippenko}
\affil{Department of Astronomy, 601 Campbell Hall, University of California,
Berkeley, CA 94720-3411, USA.} 

\author{Puragra GuhaThakurta}
\affil{UCO/Lick Observatory, Department of Astronomy \& Astrophysics, University
of California, 1156 High Street,  Santa Cruz, CA 95064, USA.}

\begin{abstract}
Supernova (SN) rates are a potentially powerful diagnostic of star formation 
history (SFH), metal enrichment, and SN physics, particularly in galaxy clusters
with their deep, metal-retaining potentials, and simple SFH. However, a
low-redshift cluster SN rate has never been published. 
We derive the SN rate in galaxy clusters at $0.06<z<0.19$, based on type~Ia
supernovae  (SNe~Ia) that were discovered by the Wise Observatory Optical
Transient Survey. As described in a separate paper, a sample of 140 rich Abell
clusters was monitored, in which six cluster SNe~Ia were found and confirmed
spectroscopically. Here, we determine the SN detection efficiencies of the
individual survey images, and combine the efficiencies with the known spectral
properties of SNe~Ia to calculate the effective visibility time of the survey.
The cluster stellar luminosities are measured from the Sloan Digital Sky Survey 
(SDSS) database in the $griz$ SDSS bands. 
Uncertainties are estimated using Monte-Carlo simulations in which all input parameters
 are allowed to vary over their known distributions.  We derive SN rates
normalized by stellar luminosity, in SNU units (SNe per century per
$10^{10}L_{\odot}$) in five photometric bandpasses, of \snuB~($B$), \snug~($g$), 
\snur~($r$), \snui~($i$), \snuz~($z$),
where the quoted errors are statistical and systematic, respectively. The SN
rate  per stellar mass unit, derived using a color-luminosity-mass relation, is 
\snuM~ SNe (century $10^{10} M_{\sun}$)$^{-1}$. The low cluster SN rates we find
are similar to, and consistent with,  the SN~Ia rate in local elliptical
galaxies. 
\end{abstract}

\keywords{ galaxies: clusters: general ---  supernovae: general --- supernovae:
individual  (SN 1998eu, SN 1998fc, SN 1999cg, SN 1999ch, SN 1999ci, SN 1999ct)}

\section{Introduction}
The event-rate history of supernovae (SNe), their progenitor systems, and the
physical  mechanisms behind these explosions, all have important implications
for several branches of astrophysics and cosmology.    As the primary sources of
iron and other heavy elements, SNe have a key role in the chemical evolution of
the Universe (e.g., Nomoto et al. 2005). The history of enrichment is determined
by the rate of SNe in the past and by the inventory of heavy elements that is
released by each explosion.    
In cosmology, type Ia SNe (SNe~Ia) used as cosmological distance indicators have
 provided direct evidence for an accelerated expansion of the Universe (e.g.,
Riess et al. 1998, 2004; Perlmutter et al. 1999; Astier et al. 2006). The use of
SNe~Ia as standard candles relies on the assumptions that there is no evolution
in their intrinsic properties, and that their luminosities can be calibrated
through an unevolving empirical relation to their light-curve shape (for
reviews, see Leibundgut 2001; Filippenko 2004, 2005). If these assumptions could 
be founded on more solid observational and physical grounds, the cosmological use 
of SNe~Ia would stand on a firmer footing.

It is widely agreed that SNe~Ia occur when a white dwarf in a binary system is
pushed  toward the Chandrasekhar mass limit and undergoes a runaway
thermonuclear explosion. However, there is no consensus regarding how this
situation is reached. In the single-degenerate scenario, a white dwarf accretes
matter from a main-sequence, subgiant, or giant star. In the double-degenerate
scenario, two white dwarfs merge. Furthermore, it is not clear how the explosion
proceeds.
Since direct observational identification of a SN~Ia progenitor system is
unlikely,  indirect evidence may be the only practical way to distinguish
between models. 
Modeling of the observed spectra of SNe provides many clues (e.g., Nomoto et al.
 1984), as do studies of Galactic SN remnants (e.g., Badenes et al.
2006). However, these studies mostly constrain the explosion physics
(deflagration or detonation) and are less sensitive to the identity of the mass
donor.

A complementary approach exploits the fact that 
each of the above scenarios may predict a ``time delay distribution,'' which is 
the distribution of times elapsed between the formation of a stellar population
and the explosions of some members of the population as SNe~Ia (e.g., Yungelson
\& Livio 2002; Greggio 2005; F\"orster et al. 2006). The time delay, in turn,
convolved with the star formation history (SFH), dictates the evolution of the
SN rate with cosmic time or with redshift. By measuring the SN~Ia rate vs.
redshift $z$, one can, in principle, constrain the form of the time delay distribution and
obtain clues about the progenitor scenario (e.g., Pain et al. 2002; Gal-Yam \&
Maoz 2004; Strolger et al. 2004; Sullivan et al. 2006; Barris \& Tonry 2006;
Neill et al. 2006; F\"orster et al. 2006). 

In practice, there are several complications. A prediction of the SN rate is
sensitive to the assumed SFH. For example, F\"orster et al. (2006) recently
re-analyzed the SN~Ia sample of Strolger et al. (2004) and argued that, by
varying the choice of SFH within the uncertainties allowed by observations, a
wide range of time delay distributions are consistent with the SN data, and
hence none of the 
SN~Ia scenarios can be ruled out. Another complication follows from the recent
evidence that SNe~Ia are descended from two distinct stellar populations
(Mannucci et al. 2005; Scannapieco \& Bildsten 2005; Sullivan et al. 2006),
perhaps leading to two SN~Ia ``channels'' -- a ``prompt'' channel, with a delay
time of order 1~Gyr or less, and a ``delayed'' channel, with delay times that
extend to $\gtrsim 10$ Gyrs. Effectively, this means a time delay distribution
that is more complex than those predicted by single-channel models.
The combination of these two issues, combined with debates regarding sample
completeness, have prevented conclusive determinations of delay times. 

Measuring the SN~Ia delay time in galaxy clusters, however, may provide a
solution to these difficulties. Analysis of the stellar population of cluster
galaxies suggests that the SFH in clusters is simpler than that in the field,
and can likely be approximated by a brief starburst at high redshift (e.g., 
Wuyts et al. 2004; Holden et al. 2005). 
Sullivan et al. (2006) show that the mix between the delayed and prompt 
components of the SN~Ia population changes in different environments. The
prompt component dominates late-type galaxies, while elliptical galaxies
contain only members of the delayed group.
SN rate measurements in clusters could then provide an estimate of the
delayed component alone with minimal contamination. If the population of
SNe~Ia is indeed bimodal, pinning down one of its components from cluster SN
studies could help to disentangle the two populations in the field.

Furthermore, measurement of cluster SN rates may help resolve the question of
the  dominant source of the high metallicity in the intracluster medium (e.g.,
Renzini et al. 1993; Maoz \& Gal-Yam 2004) -- SNe~Ia, or core-collapse SNe from
an early stellar population with a top-heavy initial mass function (e.g., Maoz
\& Gal-Yam 2004; Loewenstein 2006). Clusters are excellent laboratories for
studying enrichment, due to their simple SFH and their deep potentials from
which matter cannot escape. Therefore, understanding the source of intracluster
metals would be relevant for understanding metal enrichment in general, and the
possible role of first-generation stars in early-universe enrichment. 

Although SN~Ia rates have been measured extensively out to $z \approx 1$ (Pain et al.
 1996, 2002; Cappellaro et al. 1999; Hardin et al. 2000; Tonry et al. 2003; 
Madgwick et al. 2003; Blanc et al. 2004; Dahl\'en et al. 2004; Barris \& Tonry 2006; Neill et al. 2006), 
there have been few attempts to measure the SN rate in galaxy clusters (e.g., Barbon 1978; Norgaard-Nielsen et al. 1989; Reiss et al. 1998). 
The only published cluster SN rate at distances beyond the Virgo cluster is by 
Gal-Yam, Maoz, \& Sharon (2002). This measurement was based on the detection of
two or three likely cluster SNe~Ia in archival {\it Hubble Space Telescope} ({\it
HST}) images of high-redshift clusters. The derived rates,
$0.39^{+0.59}_{-0.25}$ SNu$_B$ at $z=0.25$ and $0.80^{+0.92}_{-0.40}$ SNu$_B$ at
$z=0.9$ [$1$ SNu$_B$ $= 1$ SN century$^{-1} (10^{10}L_{B,\sun})^{-1}]$, have
large uncertainties due to the small number of detected SNe and the lack of follow-up
observations, a consequence of searching archival data.

We are carrying out a program to measure the cluster SN rate
at both low and high redshifts. 
In this paper, we derive the cluster SN rate based on a low-redshift
($0.06<z<0.19$) cluster SN survey, the Wise Observatory Optical Transient Search
(WOOTS). 

The cluster sample, the observational design, and the SN
detection,  spectroscopic follow-up observations, and classification are described in a
separate paper by Gal-Yam et al. (2006), hereafter Paper I.

Here, we present the details of the derivation of the cluster SN rate and
briefly discuss its implications. 

Throughout the paper we assume a flat cosmology, with parameters
$\Omega_{\Lambda}=0.7$, $\Omega_{\rm M}=0.3$, and $H_0=70~{\rm km~s}^{-1} {\rm Mpc}^{-1}$. 

\section{The Survey}
In Paper I, we present the observational details of 
WOOTS, the SNe that were discovered by it, and their
followup observations.
Briefly, 
WOOTS was a survey for SNe and other transient or variable sources
in the fields of 161 galaxy clusters, and was conducted in 1997-1999.
The cluster sample was selected from the catalog 
of Abell, Corwin \& Olowin (1984),
based on the following criteria:
cluster redshift in the range $0.06<z<0.2$ (based on the compilation of Struble \&
Rood 1991); declination $\delta>0$; Abell richness class $R\geq1$ (Abell galaxy count
$N>65$); and cluster radius (estimated by Leir \& Van den Bergh 1977) smaller than $20''$.
Monthly dark-time observations used the 
Wise Observatory 1 m telescope, with a $1024\times1024$, $0\farcs7$ pixel, Tektronics CCD
imager, giving a $12'\times12'$ field of view. Images were unfiltered (see also \S~\ref{s.calib})
reaching a limiting magnitude equivalent to $R\sim 22$.
Images (of $\sim40$ clusters
per night) were compared with previously obtained template images using image subtraction
methods, and searched for transients. 

All candidate transients were followed up photometrically at Wise, 
and spectroscopically with larger telescopes. 
The determination of SN redshifts, ages and types was done by comparing the observed spectra 
to redshifted versions of high S/N template spectra of nearby SNe, 
drawn from the spectroscopic archive presented by Poznanski et al. (2002).
In the course of the survey, 12 SNe were discovered, all of them
spectroscopically confirmed (11 as type-Ia, and one as a type-IIP SN). 
Seven of the SNe (all of them SNe Ia) were in their respective
clusters, and the remaining five were foreground and background events. 
All additional transient/variable candidates were confirmed as non-SN
events: asteroids, active galactic nuclei, and variable stars.

The SN sample we will analyze in this paper
does not include SN 2001al (Gal Yam et al. 2003), a cluster SN~Ia
which was discovered during an extension of WOOTS using a new
CCD camera during its commissioning phase. 
We do not consider in this paper any of the survey images
from that survey extension. In addition, we do not include in our analysis a sub-sample 
of 21 clusters for which the imaging data were inadvertently lost.
Our sample thus consists of six confirmed cluster SNe~Ia that were discovered by
monitoring, effectively, $140$ Abell clusters at $0.06<z<0.19$.
Table~\ref{table.SNe} lists the six events and their main parameters. 

\begin{table}[ht]
\begin{center}
\caption{WOOTS Cluster SNe~Ia\label{table.SNe}}
\begin{tabular}{lll}
\tableline\tableline
SN & Cluster & Redshift\\
\tableline
1998fc&  Abell 403 & 0.10\\
1998eu&  Abell 125 & 0.18\\
1999cg&  Abell 1607& 0.14\\
1999ch&  Abell 2235& 0.15\\
1999ci&  Abell 1984& 0.12\\
1999ct&  Abell 1697& 0.18\\
\tableline\tableline		    
\end{tabular}
\end{center}
\end{table}

\section{Photometric Calibration}\label{s.calib}
A derivation of the SN rate from our survey requires knowledge
of the depth of all the survey images, and hence a photometric
calibration of the survey data in the observed bandpass.
The WOOTS images were obtained in
unfiltered (``clear'') mode, resulting in a very broad, 
nonstandard photometric bandpass.
We have determined the effective WOOTS-clear bandpass by multiplying
the quantum
efficiency curve of the back-illuminated, Lumogen-coated, Tektronix
CCD used in the survey,
by the typical atmospheric transmission at Wise Observatory for the 
mean airmass (1.2) of the WOOTS observations. We have verified that
the effects of varying airmass and of mirror reflectivity
vs. wavelength have negligible effect of the shape of the bandpass, The
peak transmission of the effective WOOTS-clear band 
is at $\sim 7000$~\AA, with a
full-width at half maximum (FWHM) of $\sim 4700$~\AA. 

To obtain the individual photometric 
zeropoints of all the WOOTS images, we begin by using
those images that fall within the coverage of the Sloan Digital Sky Survey Data
Release 4 (SDSS DR4; Adelman-McCarthy et al. 2006). We extract the SDSS
photometry of objects identified in each image and find the best fit to the
SDSS $ugriz$ magnitudes from among a set of template spectra of stars (with
spectral types O to M; Silva et al. 1992) and galaxies (Kinney et al. 1996). 
We then use the WOOTS-clear bandpass to perform synthetic, Vega-based
photometry on the best-fit spectrum for each object. Finally,
we compare these synthetic 
magnitudes to the instrumental magnitudes of the same objects, defined as
$m_{\rm inst}=-2.5\log({\rm counts})$.
The WOOTS zeropoint is then defined as the median of the differences between
the synthetic and the instrumental magnitudes of the calibration sources
in the image.
The root-mean-square (rms) scatter
of these differences defines the statistical zeropoint error for each image. 
Every source in an image can then be calibrated to a Vega-based
WOOTS-clear magnitude by adding to its instrumental magnitude
the derived zeropoint of the image.
We estimate the accuracy of the zeropoints measured this way by adding the
systematic uncertainty in the SDSS zeropoints ($0.02$ mag) to the statistical
zeropoint error determined for each image, as defined above (typically also
$0.02$ mag), resulting in a total zeropoint error of $0.04$ mag. 

In order to calibrate the images that are in areas not covered by the SDSS, we
compare the instrumental magnitudes of objects in our images with their ``$R1$''
magnitudes in the USNO-$B$ catalog (Monet et al. 2003), to get an ``$R1$''-based
zeropoint. The USNO-$R1$ band has a similar effective wavelength to that of the
WOOTS-clear band, but is much narrower. From $329$ images in regions covered by
the SDSS, we measure a mean offset of $0.24$ mag between the Vega-based
unfiltered WOOTS zeropoint and the $R1$ zeropoint, and use it to calibrate the
rest of the images. The scatter around this offset, $0.3$ mag, reflects both
image-to-image scatter in the calibration process (e.g., due to variations in
the mix of spectral types among the lists of calibration objects), and
systematic errors in the USNO-$B$ photometry. The photometric accuracy of USNO-$B$
is not well determined, but \citet{2003AJ....125..984M} estimate that the
scatter in the photometric solution over the entire catalog has an rms of $0.25$
mag, consistent with our results. 
We therefore adopt $0.3$ mag as the zeropoint error of the non-SDSS images.
These calibration uncertainties affect the overall uncertainty in the SN rate, 
through the determination of our search efficiency and the visibility time 
(see \S~\ref{S.results}).

\section{SN Rate Calculation}
 
Given a sample of $N$ SNe discovered as part of a survey, 
the SN rate per unit stellar luminosity is
\begin{equation}
{\cal R}_{\rm Ia} = \frac{N}{\sum\limits_{j}{\Delta t_j L_{band,j}}},
\end{equation}\label{eq.SNR}
\noindent where $\Delta t_j$ is the effective visibility time (or ``control
time''), i.e., the time during which a cluster SN~Ia is above the
detection limit of the $j$th image, $L_{band, j}$ is the cluster luminosity
within the search area in a chosen photometric band, and the summation is over
all the survey images\footnote{By ``image'' we  will refer to the summed
exposures of a field from a given night.}. 
We explain below our measurement of each of these variables in the present
survey.

\subsection{Visibility Time and Light Curves}
The effective visibility time, $\Delta t_j$, can be understood as the amount of
time during which the survey is sensitive to SNe in a comparison between two
particular images of a given cluster. 
This quantity depends on the SN detection efficiency in each image, the SN light
curve at each cluster redshift, and on the frequency of searches in the same
field. 
We calculate the effective visibility time from
\begin{equation}\label{eq_deltat}
\Delta t_j = \int^{\infty}_{-\infty} {\eta^*[m_{eff}(t)]dt},
\end{equation}
where $m_{eff}(t)$ is an ``effective'' SN~Ia light curve in the survey bandpass
at the given redshift, as explained in \S~\ref{S.LC} below, and $\eta
^*[m_{eff}(t)]$ is the detection probability as a function of SN effective
magnitude. 
We describe below each step in this calculation.  

\subsubsection{Detection Efficiency Estimation}\label{S.efficiency}
\begin{figure*}[ht]
\plotone{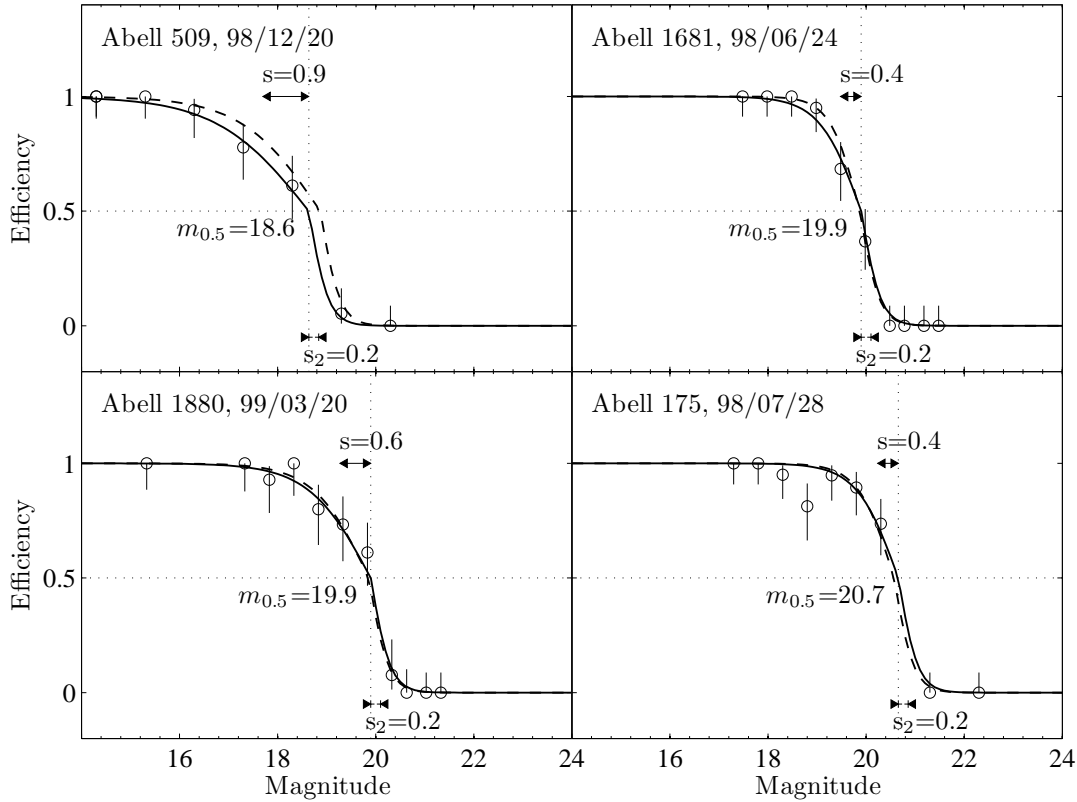}
\caption{Examples of point-source detection efficiency curves, 
as a function of WOOTS-clear Vega-based magnitude,
for four survey   images. Cluster names and observation dates are as marked.
Circles mark the fraction of detected fake SNe, with error bars based on a
binomial distribution. Solid curves are the best-fit efficiency curves
(Eq.~\ref{eq.efficiency}), and the dashed curves are the efficiency curves that
correspond to the observational parameters of each image and the relations in
Fig.~\ref{fig_correlations}. \label{fig.efficiency}} 
\end{figure*}

The detection efficiency of an image as a function of magnitude is the
probability of detecting a SN-like point source in that image.  
In order to estimate the detection efficiency for a given field, we carry out
simulations, following the prescription in Gal-Yam et al. (2002). About $200$
fake SNe are added blindly to each field, with a range of magnitudes, and with a
spatial distribution that follows the flux of galaxies in the field. The
simulated data undergo the same search procedure as the real data by means of
difference image analysis (see Paper I), and the number of fake SNe that are
successfully detected in each magnitude bin is noted.  
The SN detection efficiency is usually close to 100\% 1--2 mag above the
detection limit of the image, and drops roughly linearly to zero over this range
(Fig. \ref{fig.efficiency}).  
We parametrize the efficiency curve with the function 
\begin{equation}\label{eq.efficiency}
\eta(m;m_{0.5},s,s_2) = \left\{ \begin{array}{ll}
      \left(1 + e^{\frac{m-m_{0.5}}{s}}\right)^{-1}, & \mbox{$m\le m_{0.5}$}\\
      \left(1 + e^{\frac{m-m_{0.5}}{s_2}}\right)^{-1}, & \mbox{$m>m_{0.5}$},\\
      \end{array} \right.
\end{equation}
where $m$ is the Vega-based magnitude of a SN in the effective bandpass of the
survey (see below), $m_{0.5}$ is the magnitude at which the efficiency drops to
$0.5$, and $s$ and $s_2$ determine the range of $m$ over which $\eta$ changes
from $1$ to $0.5$ and from $0.5$ to zero, respectively.  
The main contribution to a slow convergence to unity at magnitudes brighter than
$m_{0.5}$ is the difficulty of detecting SNe which lie close to, or are superposed on,
the nucleus of a bright galaxy. Such cases are automatically included in our
simulations when distributing the fake SNe in the images. 
Figure \ref{fig.efficiency} shows four examples of the results of efficiency
simulations, and the best-fit efficiency curves. 

The WOOTS database consists of $913$ images, which were obtained under a range
of observing conditions -- atmospheric transparency, mirror reflectivity, sky
background, and telescope image quality. As a result, the detection efficiency
varies significantly among the images. Since performing efficiency simulations in
each and every image would be  impractical, we have carried out simulations, as
described above, for a subset of 30 images spanning a range in observing
conditions. We then searched for correlations between the parameters of the
resulting efficiency curves and various parameters that characterize each
image. 
As detailed below, we found that the detection efficiency is primarily determined by two
image parameters: 
the photometric zeropoint, which determines $m_{0.5}$, and the
number of residuals that are left in each difference image, with which the upper
slope parameter, $s$, is correlated. The lower slope parameter, $s_2$, did not
vary significantly among the 30 simulated images, and was therefore set to its
mean value, $0.2$. 

\begin{figure*}[ht]
\plotone {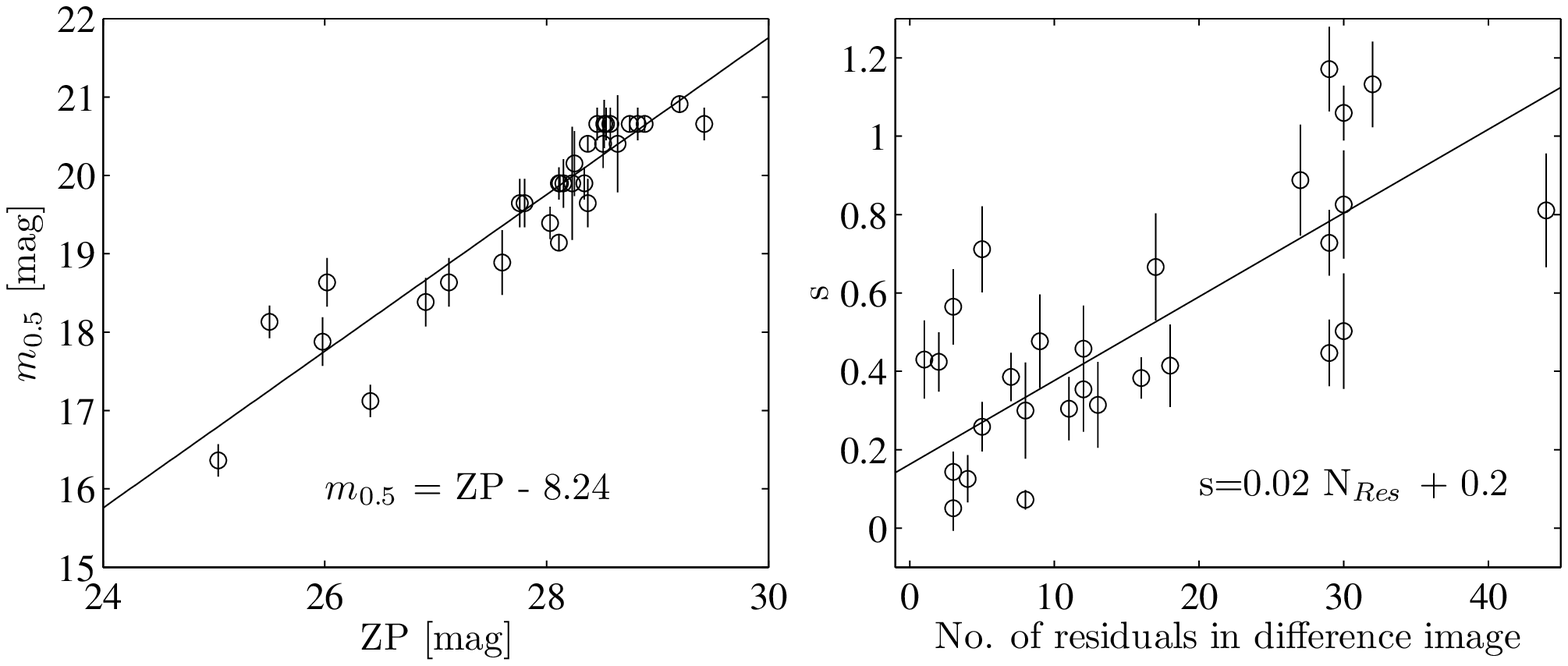}
\caption{Correlations between efficiency function parameters and image properties. 
{\it Left}: Magnitude at half the maximum efficiency ($m_{0.5}$)
vs. image zeropoint ($ZP$), both in WOOTS-clear Vega magnitudes. 
{\it Right}: Exponential decline rate ($s$) vs. number of residuals in the difference image ($N_{Res}$). 
\label{fig_correlations}}
\end{figure*}

The sensitivity of each image to SNe is determined by its photometric zeropoint,
which combines the effects of detector quantum efficiency, variable 
mirror reflectivity,
and variable atmospheric transparency. Specifically,
the zeropoint, $ZP$, 
is an indicator of the limiting magnitude of the image, and
therefore dictates the magnitude around which our detection efficiency drops
rapidly. Figure~\ref{fig_correlations} shows the relation that we have found
between $ZP$ and $m_{0.5}$, $m_{0.5}=ZP-8.25$, with a scatter of $0.42$ mag. 

The number of residuals, $N_{Res}$, was defined as the
number of objects detected by SExtractor (Bertin \& Arnouts 1996) in each difference image, with a
detection threshold of $5\sigma$ above the background.  A large number of
subtraction residuals in an image can be due to several reasons. Most often,
poor subtraction occurs when the point-spread function (PSF) in the image is very different from the PSF
in the reference image. This happens when the PSF is distorted, e.g., due to
imperfect telescope tracking, or atmospheric refraction at high airmass.
Imperfect PSF matching and poor image subtraction will mostly affect the cores
of bright galaxies, decreasing the chances of detecting SNe that lie close to
galactic nuclei, and causing the efficiency curve to converge more slowly to the
maximal detection efficiency. This behavior is  parametrized by $s$. We find
that $s$ depends on the number of residuals through the relation $s=0.02N_{Res}
+ 0.3$, with an rms of $0.3$ (see Fig. \ref{fig_correlations}). 

We found no significant dependence of detection efficiency on the remaining
observational parameters (sky background level and seeing width) 
due to the fact
that most of the images were obtained during dark time, and the seeing spanned a
limited range of $3\pm0.4''$ FWHM. 
Based on the above
relations, we can obtain efficiency curves for all images in the survey based on
their measured observational parameters, $ZP$ and $N_{Res}$.

\subsubsection{Light Curves}\label{S.LC}
The peak magnitudes of SNe~Ia exhibit an intrinsic rms scatter of 0.2--0.3 mag,
and are correlated with the shape of the light curve through a stretch relation
(Phillips 1993; see Leibundgut 2001 for a review). Since more luminous SNe tend to
rise and decline more slowly than less luminous SNe, their overall visibility time is
longer than that of dim SNe.  
Sullivan et al. (2006b) study the dependence of SN~Ia properties on host-galaxy
type, and show that SNe~Ia in elliptical galaxies tend to be dimmer (with a
smaller stretch factor). In their Fig. 11, they present the distribution of
stretch factors according to galaxy type, at low and high redshifts.  
The bulk of the stellar light and mass in clusters is contributed 
by early-type galaxies. Indeed, among the six cluster SNe in our sample, the five with hosts were in
early-type galaxies (one of the SNe was a hostless intergalactic cluster SN --
see Gal-Yam et al. 2003). 
Assuming that all the SNe~Ia in the clusters in our sample occur in
elliptical galaxies, we use the distribution of stretch factors of
SNe with elliptical hosts to assemble a dataset of light curves that will serve
as templates. 
We base these light curves on a rest-frame, non-stretched, $B$-band template
light curve from \citet{2002PASP..114..803N}, and transform it to different
stretched light curves using the stretch relation, $M_s = M_{s=1} - \alpha(s-1)$,$t_s = t_{s=1}\times \alpha$ as described by \citet{1999ApJ...517..565P}.
For consistency, we use the same non-stretched peak $B$-band magnitude and $\alpha$ as 
Sullivan et al. (2006a), $M_B = -19.25$ (for h=0.7) and $\alpha = 1.47$, which are based on the 
results of Knop et al. (2003). We assume an uncertainty of $0.15$ mag in $M_B$, 
from the dispersion in peak magnitudes of local SN light curves after stretch correction (Guy et al. 2005). 
This uncertainty is taken into account in our error budget (see \S~\ref{S.results}).

\begin{figure}[ht]
\plotone{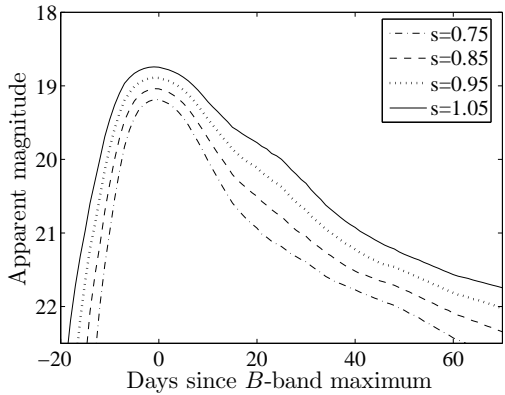}
\caption{Examples of template SN~Ia light curves with a range of stretch factors, 
calculated in the  WOOTS-clear bandpass.  In this example, the light curves are 
for the cluster Abell 1920, at $z=0.131$. 
The representative stretch values for SNe~Ia in low-$z$ elliptical galaxies are
from Sullivan et al. (2006). 
\label{fig_lightcurves}}
\end{figure}

Since WOOTS observations were unfiltered, a transformation from the  rest-frame
$B$-band light curve to an observed WOOTS-clear 
light curve at the cluster redshift is
necessary.  For each cluster, we calculate a set of stretched SN light curves
from multi-epoch SN~Ia spectra, using synthetic photometry, as follows.  
We start with a set of multi-epoch spectral templates from
\citet{2002PASP..114..803N}. For each combination of stretch and redshift, we
normalize the spectra to fit the $B$-band rest-frame photometry, and redshift
them according to the cluster redshift. We then multiply the spectra by the
WOOTS-clear bandpass, to obtain an unfiltered light curve. 
The result is a set of representative WOOTS-clear 
light curves for every cluster,
each light curve with a different stretch. 
Figure~\ref{fig_lightcurves} shows an
example of the template light curves for a cluster at $z=0.131$. 
When calculating the visibility time for a particular image, we draw for it a
stretch factor (with its corresponding, properly normalized, light curve) from
the Sullivan et al. (2006) distribution of stretch factors.    

\subsubsection{Detection Probability}

The detection efficiency function described in \S~\ref{S.efficiency} is defined for
the magnitude of a point source in a {\it difference} image. This means that
the interesting quantity in the SN light curve is not its magnitude in a single
image, but the magnitude corresponding to the flux difference $\Delta f$ that
results from subtracting two epochs from each other. This effective light curve
depends on the time that elapsed between the two epochs that are being
compared, 
\begin{eqnarray}\label{eq.meff}
m_{eff}(t)&=&ZP-2.5\log(\Delta f)=\\\nonumber
&&-2.5\log\left(10^{-0.4m(t)}-10^{-0.4m[t+(t_1-t_0)]} \right),
\end{eqnarray}
where $t$ is the age of the SN, and $t_0$ and $t_1$ are the times of the
template observation and the new observation, respectively. 

The detection probability function, $\eta[m_{eff}(t)]$, describes the probability
of detecting a SN which occurred $t$ days ago, in a comparison between two epochs.
 
Since the searches in a given field were not uniformly distributed in time, with
some observations conducted close in time to each other, the same SN can, in
principle, be discovered in more than one image.  
For example, when two closely spaced observations were compared to the same
template, a SN that was discovered in the first observation (at time $t_1$)
would simply be rediscovered in the second observation (at time $t_2$). 
To avoid counting such objects twice in our calculation of the visibility time,
we assign the second observation a reduced detection efficiency, which describes
the probability of detecting only a SN that was  {\it not} detected in the previous
observation. 
We do this by shifting $\eta(t) $ by $t_2-t_1$ days, and subtracting from it the
combined detection probability (the probability that the SN is discovered in
both observations): $\eta _2^* = \eta _2 - (\eta _2 \times \eta _1$). 
In principle, this operation can be repeated for any number of consecutive observations, e.g.,
for three consecutive observations, $\eta _3^*$ will be
the probability to discover a SN in the third observation ($\eta _3$), 
minus the probabilities of discovering it
both in the first observation and in the third one, the second and the third,
and in all three observations: 
$\eta _3^* = \eta _3 - (\eta _3 \times \eta _2) - (\eta _3 \times \eta _1) - (\eta _3
\times \eta _2 \times \eta _1$). 
In practice, the temporal distribution of the images does not require
considering more than two consecutive observations, i.e., either $\eta^*=\eta
_1$ or $\eta^*=\eta _2^*$ are used in equation~\ref{eq_deltat}.

\subsection{Cluster Stellar Luminosity}\label{s.luminosity}
\subsection{Aperture Luminosity Measurement}\label{s.luminosity1}
SN rates are often measured relative to the stellar luminosity within 
the search
area, in a particular band. The cluster luminosities in our sample cannot be
measured easily from WOOTS data, since they lack color and spectral
information from which cluster membership could be determined.  
Instead, we have used the data from the SDSS DR4 to measure cluster luminosities
for 72 clusters, i.e., about one-half of the sample. 
Unlike traditional cluster stellar luminosity measurements, we do not identify
the individual cluster member galaxies from the data. Instead, we have based our
measurements on a method similar to aperture photometry, as follows.  

The net cluster-galaxy flux is
\begin{equation}
f_{cluster} = k \times ( \sum_{i}f_i - \pi R_c^2 \overline{f}_{bg}),
\end{equation}
where $k$ is the K-correction factor, and the summation is over all galaxies
within a cluster radius $R_c$ that satisfy the criteria that will be described
below.  The total Galactic extinction-corrected flux\footnote{Fluxes are based
on the ``modelmag'' SDSS magnitudes.} of galaxy $i$ is denoted by $f_i$, and 
$\overline{f}_{bg}$ is
the average ``background''-galaxy\footnote{Non-cluster galaxies are, of course, both
foreground and background, but we retain the term in analogy to aperture
photometry.} 
flux per unit area. 
The net cluster-galaxy flux is translated to luminosity based on the cluster
redshift and the appropriate K-correction, 
and corrected for incompleteness, as described below. In what
follows, we describe the derivation of cluster stellar luminosities in the
SDSS $r$ band. The luminosities in three other SDSS photometric bands, $g,i,$
and $z$, were measured in the same manner.  

To lessen the contamination of our measurements by foreground galaxies, we
ignored, both within and outside our apertures,
 galaxies brighter than the brightest-cluster galaxy (BCG) in each field.
The BCG was identified from the SDSS catalog by its magnitude, color, and where
available, by redshift. We also ignored galaxies fainter than the SDSS $95\%$
completeness limit (see Table \ref{table.parameters}). 

\begin{table}[ht]
\begin{flushleft}
\caption{Adopted Photometric Parameters\label{table.parameters}}
\begin{tabular}{ccccc}
\tableline\tableline
	&(1)       &(2)     &(3)                &(4)     \\
band	&$m_{lim}$&$M_{{\rm band}, \odot}$&$M^*$  &$\alpha$ \\        
	&[mag]    &[mag]    &[mag]             &          \\
\tableline	
$g$	&$22.2$ & $5.12$& $-22.01\pm0.26$& $-1.00\pm0.06$ \\   
$r$	&$22.2$ & $4.64$& $-22.21\pm0.05$& $-0.85\pm0.03$ \\
$i$	&$21.3$ & $4.53$& $-22.31\pm0.08$& $-0.70\pm0.05$ \\
$z$	&$20.5$ & $4.51$& $-22.36\pm0.06$& $-0.58\pm0.04$ \\
\tableline\tableline		
\end{tabular}\\
\tablecomments{(1) SDSS 95\% completeness magnitude; Adelman-McCarthy et al. (2006).\\
(2) Solar absolute magnitudes; Blanton et al. (2006).\\
(3),(4) Schechter luminosity function parameters for clusters; Goto et al. (2002).}
\end{flushleft}
\end{table}

Leir \& van den Bergh (1977) estimated cluster radii, 
$R_{LvdB}$, for 1889 Abell
clusters, using the red plates of the Palomar Sky Survey, by superposing a grid
of circles of various radii on each cluster, and selecting by eye the radius
that encompasses most uniformly all visible cluster members. We have examined
their estimate for 12 of our clusters, by measuring the net cluster-galaxy flux
as described above, as a function of $R_c$. We find that the enclosed flux
profile generally tends to a constant value (to within errors) at radii
$R_c\approx R_{LvdB}$, and we therefore adopt $R_{LvdB}$ as the cluster radius
of each cluster in our sample. In \S3 we investigate the effect of varying the
choice of $R_c$ between $0.5 R_{LvdB}$ and $1.5 R_{LvdB}$, according to a normal distribution centered on $R_{LvdB}$ with $\sigma=0.2R_{LvdB}$. We find that this uncertainty in the cluster radius results in a $\sim 1\%$ error in the stellar luminosity.

The average background-galaxy flux per unit area, $\overline{f}_{bg}$, was
estimated from the flux per unit area of galaxies that satisfy the above
criteria, in an annulus of area of $\sim0.5$ deg$^2$ centered on the cluster,
with inner radius of $0.4^\circ$. The inner radius is chosen to be large enough
to exclude cluster member galaxies, but small enough to be 
representative of the
large-scale structure in the vicinity of the cluster. 
In order to check whether the measured cluster luminosity is sensitive to the
choice of background annulus inner radius, we calculated the luminosity of each
cluster using 
 a range of inner radii, between $0.35^\circ$ and $0.45^\circ$. We
experimented with both using a fixed 
angular radius for all of the clusters, and
using a buffer that is proportional to the cluster radius. The luminosities
that result from the various methods of background estimation have an rms of $8\%$. We take this uncertainty into account in our error budget
(see \S\ref{S.results}). 

To convert the observed magnitudes to rest-frame magnitudes, a K-correction
factor was calculated for each cluster assuming that all the cluster light is
emitted by elliptical galaxies at the cluster redshift.
The net cluster flux
was then translated to luminosity, according to the cluster redshift and the adopted
cosmology. 
The faint cut on the galaxy magnitudes means that the summed luminosity includes
only galaxies brighter than some limiting luminosity, which depends on the
cluster redshift. We correct for this incompleteness by multiplying the summed
luminosity by the fraction of light that comes from the faint end of a Schechter
(1976) luminosity function: 
\begin{equation}
C = \frac{\int _0^\infty {\Phi (L)dL} } { \int _{L_{lim}(m_{lim},z)}^\infty \Phi(L)dL},
\end{equation}
where 
$\Phi(L) dL = \Phi^*(L/L^*)^\alpha {\rm exp}(-L/L^*)d(L/L^*)$. We adopt
$\alpha=-0.85\pm 0.03$ and $M^*=-22.21 \pm0.05$ as the mean values for the
$r$-band luminosity function parameters in clusters (Goto et al. 2002; see
Table~\ref{table.parameters}). 
The integrated $r$-band luminosity correction
factors in our sample are in the range  $C-1 = 0.4^{+0.09}_{-0.07}\times 10^{-3}$ to
$8.3^{+1.1}_{-0.9}\times 10^{-3}$, for $z=0.06$ and $z=0.2$, respectively,
i.e., always quite small (see Table \ref{table.clusters}).   
We note that Goto et al. (2002) excluded the BCGs from the analysis
when fitting the data to the Schechter function. We therefore applied the correction above 
after subtracting the luminosity of the BCG from the summed luminosity, and added the BCG luminosity 
to the corrected cluster luminosity.

Several recent studies have argued that BCG galaxy
luminosities are underestimated in SDSS,
because the default sky subtraction algorithm
removes the outer, low surface-brightness,
flux from these galaxies (Graham et al. 2005;
Bernardi et al. 2005; Lauer et al. 2006; Desroches et al. 2006;
von der Linden et al. 2006). L.-B. Desroches
and C.-P. Ma (private communication) 
find that, for a subsample of the Miller et al. (2005) C4 sample,
improved sky subtraction increases the luminosity by a mean factor 1.3
over the total deVaucoulers SDSS luminosity. For five of our BGCs
which are in common with their subsample they find similar
correction factors. It is clear that BCG luminosity measurements
from the SDSS are indeed sensitive to the sky-level subtraction
algorithm. On the other hand, it is not certain that the new
corrected measurements are in fact superior, e.g., they might
include in the BCG measurement some light from neighboring
galaxies. We therefore choose to increase the BCG luminosities
by 15\%, with an additional systematic uncertainty of $\pm 15\%$.
Since the BCG luminosity typically constitutes about 6\% of
the cluster luminosity, this correction lowers our derived SN rate
by 1\%, and increases the systematic error in the rate by $\pm 1\%$.

\begin{table}[ht]
\begin{flushleft}
\tiny
\caption{Measured Cluster Luminosities\label{table.clusters}}
\begin{tabular}{llllllll}
\tableline\tableline
Abell	& $z$ 	& \multicolumn{4}{c}{Luminosity [$10^{12}L_{\sun}$]} & Luminosity & Mass\\ 
Number  &  	& $g$ 	& $r$ 	& $i$ 	& $z$ 	& Correction (r) &  [$10^{12}M_{\sun}$]\\ 
\tableline 
125	& 0.19	& 2.51	& 2.51	& 3.10	& 4.50	& 1.0071	& 8.2 $\pm 1.9$ \\ 
 175	& 0.13	& 4.67	& 6.03	& 7.59	& 8.52	& 1.0027	& 16.3 $\pm 3.8$ \\ 
 279	& 0.08	& 0.82	& 0.91	& 1.41	& 1.61	& 1.0008	& 5.0 $\pm 1.2$ \\ 
 655	& 0.12	& 4.55	& 5.00	& 6.99	& 8.14	& 1.0025	& 19.8 $\pm 4.7$ \\ 
 917	& 0.13	& 1.04	& 1.40	& 1.79	& 2.15	& 1.0029	& 4.2 $\pm 1.0$ \\ 
 924	& 0.10	& 0.60	& 0.83	& 1.13	& 1.29	& 1.0014	& 2.9 $\pm 0.7$ \\ 
 947	& 0.18	& 1.55	& 2.06	& 2.32	& 2.27	& 1.0061	& 3.3 $\pm 0.8$ \\ 
 975	& 0.12	& 0.87	& 0.94	& 1.20	& 1.66	& 1.0022	& 3.3 $\pm 0.8$ \\ 
 1025	& 0.15	& 2.81	& 3.30	& 3.77	& 4.77	& 1.0041	& 7.3 $\pm 1.7$ \\ 
 1066	& 0.07	& 1.60	& 1.86	& 2.26	& 2.71	& 1.0006	& 4.8 $\pm 1.1$ \\ 
 1073	& 0.14	& 2.27	& 3.18	& 3.46	& 4.56	& 1.0033	& 6.2 $\pm 1.5$ \\ 
 1081	& 0.16	& 3.41	& 4.02	& 5.49	& 6.23	& 1.0046	& 14.4 $\pm 3.4$ \\ 
 1132	& 0.14	& 3.74	& 5.16	& 6.58	& 7.53	& 1.0031	& 14.8 $\pm 3.5$ \\ 
 1170	& 0.16	& 3.83	& 3.54	& 4.29	& 5.04	& 1.0049	& 8.8 $\pm 2.1$ \\ 
 1190	& 0.08	& 1.25	& 1.54	& 2.18	& 2.70	& 1.0008	& 6.8 $\pm 1.6$ \\ 
 1201	& 0.17	& 7.05	& 8.61	& 9.57	& 11.26	& 1.0054	& 16.1 $\pm 3.8$ \\ 
 1207	& 0.14	& 0.80	& 0.71	& 0.96	& 1.59	& 1.0031	& 3.5 $\pm 0.8$ \\ 
 1227	& 0.11	& 2.45	& 3.19	& 4.50	& 4.65	& 1.0019	& 11.5 $\pm 2.7$ \\ 
 1474	& 0.08	& 1.17	& 1.52	& 1.84	& 2.12	& 1.0008	& 3.7 $\pm 0.9$ \\ 
 1477	& 0.11	& 0.08	& 0.15	& 0.18	& 0.24	& 1.0019	& 0.4 $\pm 0.1$ \\ 
 1524	& 0.14	& 1.77	& 2.07	& 2.50	& 3.12	& 1.0032	& 5.4 $\pm 1.3$ \\ 
 1528	& 0.15	& 1.34	& 1.78	& 2.44	& 3.22	& 1.0043	& 7.5 $\pm 1.8$ \\ 
 1539	& 0.17	& 3.15	& 4.32	& 5.23	& 6.76	& 1.0056	& 11.7 $\pm 2.8$ \\ 
 1552	& 0.08	& 1.95	& 2.11	& 2.68	& 3.30	& 1.0010	& 6.4 $\pm 1.5$ \\ 
 1553	& 0.17	& 5.46	& 5.39	& 6.83	& 8.30	& 1.0051	& 16.1 $\pm 3.8$ \\ 
 1566	& 0.10	& 0.75	& 0.84	& 1.13	& 1.44	& 1.0015	& 3.2 $\pm 0.8$ \\ 
 1617	& 0.15	& 2.60	& 2.77	& 3.58	& 4.62	& 1.0041	& 9.3 $\pm 2.2$ \\ 
 1661	& 0.17	& 1.44	& 2.15	& 2.87	& 3.36	& 1.0053	& 7.4 $\pm 1.7$ \\ 
 1667	& 0.17	& 2.15	& 2.38	& 2.92	& 3.69	& 1.0051	& 6.6 $\pm 1.6$ \\ 
 1674	& 0.11	& 1.56	& 1.95	& 2.78	& 2.95	& 1.0017	& 7.5 $\pm 1.8$ \\ 
 1677	& 0.18	& 2.36	& 2.82	& 3.33	& 4.17	& 1.0066	& 6.9 $\pm 1.6$ \\ 
 1678	& 0.17	& 1.88	& 2.32	& 2.71	& 3.23	& 1.0055	& 5.2 $\pm 1.2$ \\ 
 1679	& 0.17	& 1.89	& 2.26	& 2.67	& 3.32	& 1.0055	& 5.4 $\pm 1.3$ \\ 
 1697	& 0.18	& 3.20	& 3.82	& 5.14	& 5.96	& 1.0066	& 13.2 $\pm 3.1$ \\ 
 1731	& 0.19	& 2.42	& 2.87	& 3.78	& 5.16	& 1.0076	& 10.9 $\pm 2.6$ \\ 
 1738	& 0.12	& 2.27	& 2.91	& 3.41	& 4.83	& 1.0021	& 7.8 $\pm 1.8$ \\ 
 1763	& 0.19	& 4.01	& 5.19	& 6.60	& 7.63	& 1.0070	& 14.9 $\pm 3.5$ \\ 
 1767	& 0.07	& 1.28	& 1.60	& 1.95	& 2.70	& 1.0006	& 4.8 $\pm 1.1$ \\ 
 1773	& 0.08	& 1.35	& 1.57	& 2.00	& 2.25	& 1.0008	& 4.4 $\pm 1.0$ \\ 
 1774	& 0.17	& 2.46	& 2.85	& 3.23	& 4.65	& 1.0054	& 7.0 $\pm 1.6$ \\ 
 1780	& 0.08	& 1.39	& 1.69	& 2.16	& 2.31	& 1.0008	& 4.6 $\pm 1.1$ \\ 
 1795	& 0.06	& 1.56	& 1.80	& 2.19	& 3.48	& 1.0005	& 6.1 $\pm 1.4$ \\ 
 1889	& 0.19	& 3.11	& 3.04	& 4.02	& 5.06	& 1.0069	& 10.8 $\pm 2.5$ \\ 
 1911	& 0.19	& 2.71	& 2.69	& 3.43	& 4.04	& 1.0074	& 8.0 $\pm 1.9$ \\ 
 1914	& 0.17	& 6.01	& 7.08	& 8.52	& 9.98	& 1.0056	& 17.1 $\pm 4.0$ \\ 
 1918	& 0.14	& 1.45	& 1.87	& 2.25	& 2.68	& 1.0034	& 4.6 $\pm 1.1$ \\ 
 1920	& 0.13	& 2.55	& 2.72	& 3.47	& 4.68	& 1.0029	& 9.2 $\pm 2.2$ \\ 
 1926	& 0.13	& 4.46	& 5.05	& 6.41	& 7.98	& 1.0029	& 15.5 $\pm 3.6$ \\ 
 1936	& 0.14	& 1.42	& 1.68	& 2.12	& 2.51	& 1.0033	& 4.8 $\pm 1.1$ \\ 
 1937	& 0.14	& 0.63	& 0.89	& 1.23	& 1.80	& 1.0033	& 4.2 $\pm 1.0$ \\ 
 1940	& 0.14	& 3.67	& 4.13	& 4.89	& 6.30	& 1.0034	& 10.4 $\pm 2.5$ \\ 
 1954	& 0.18	& 2.31	& 2.56	& 3.09	& 3.73	& 1.0064	& 6.5 $\pm 1.5$ \\ 
 1966	& 0.15	& 0.84	& 1.21	& 1.53	& 3.09	& 1.0041	& 5.9 $\pm 1.4$ \\ 
 1979	& 0.17	& 2.24	& 2.58	& 3.10	& 3.54	& 1.0054	& 6.1 $\pm 1.4$ \\ 
 1984	& 0.12	& 1.24	& 1.54	& 1.97	& 2.32	& 1.0025	& 4.6 $\pm 1.1$ \\ 
 1986	& 0.12	& 1.33	& 1.78	& 1.97	& 2.52	& 1.0022	& 3.6 $\pm 0.8$ \\ 
 1990	& 0.13	& 0.89	& 1.40	& 1.98	& 2.28	& 1.0026	& 5.6 $\pm 1.3$ \\ 
 1999	& 0.10	& 2.61	& 2.98	& 3.88	& 4.70	& 1.0016	& 9.7 $\pm 2.3$ \\ 
 2005	& 0.13	& 2.21	& 2.79	& 3.47	& 4.01	& 1.0026	& 7.4 $\pm 1.7$ \\ 
 2008	& 0.18	& 1.28	& 2.20	& 2.78	& 3.74	& 1.0064	& 7.2 $\pm 1.7$ \\ 
 2009	& 0.15	& 2.73	& 3.72	& 4.40	& 6.00	& 1.0042	& 9.9 $\pm 2.3$ \\ 
 2029	& 0.08	& 2.49	& 2.85	& 3.36	& 5.62	& 1.0008	& 9.2 $\pm 2.2$ \\ 
 2061	& 0.08	& 1.72	& 1.64	& 2.17	& 2.09	& 1.0008	& 4.4 $\pm 1.0$ \\ 
 2062	& 0.11	& 1.99	& 2.25	& 2.94	& 3.56	& 1.0019	& 7.4 $\pm 1.8$ \\ 
 2089	& 0.07	& 0.51	& 0.79	& 0.94	& 0.84	& 1.0007	& 1.4 $\pm 0.3$ \\ 
 2100	& 0.15	& 3.03	& 3.80	& 4.69	& 5.78	& 1.0042	& 10.5 $\pm 2.5$ \\ 
 2122	& 0.07	& 0.68	& 0.91	& 1.13	& 1.44	& 1.0005	& 2.7 $\pm 0.6$ \\ 
 2172	& 0.14	& 1.21	& 1.59	& 1.97	& 2.55	& 1.0033	& 4.7 $\pm 1.1$ \\ 
 2213	& 0.16	& 1.15	& 1.39	& 1.52	& 2.05	& 1.0047	& 2.8 $\pm 0.7$ \\ 
 2235	& 0.15	& 1.45	& 1.78	& 2.27	& 2.64	& 1.0041	& 5.2 $\pm 1.2$ \\ 
 2244	& 0.10	& 1.83	& 2.76	& 3.38	& 3.31	& 1.0014	& 5.9 $\pm 1.4$ \\ 
 2255	& 0.08	& 3.64	& 4.14	& 5.46	& 6.00	& 1.0009	& 12.7 $\pm 3.0$ \\ 
 \tableline		    
\end{tabular}
\\
\end{flushleft}
\end{table}

\subsubsection{Comparison to Other Luminosity Measurements}
The reliability of our luminosity measurement method can be tested by comparing
our results to  independent measurements of the same clusters. The comparison
can be particularly straightforward, if the compared luminosities were derived
from the same data, namely, the SDSS.  
\citet{2005AJ....130..968M} describe their compilation of a cluster catalog,
containing $748$ clusters selected spectroscopically from the SDSS database, 12
of which are in our sample and are fully covered by the SDSS DR4.  
\citet{2005AJ....130..968M} calculate the $r$-band luminosities of the clusters
as the summed luminosities of all galaxies within $2.1$ Mpc (for our choice of
$H_0$) from the center of the cluster, within 4$\sigma$ in redshift space, and
more luminous than $1.4\times 10^{10} L_{\sun}$ (corresponding to a $17.8$ mag
galaxy at $z=0.11$).  
In order to compare our results to theirs, we have calculated the aperture-based
luminosities for the $12$ clusters that appear in both samples, with the same
radius and limiting luminosity criteria. Excluding
Abell 1539, for which the luminosity measurement of
Miller et al. was corrupted by a satellite trail in the
SDSS $r$-band image, we find
that the mean difference between the two
measurements of every cluster is $\sim20\%$. 

\subsubsection{Measured Luminosities}\label{S.measured}
\begin{figure}[ht]
\plotone {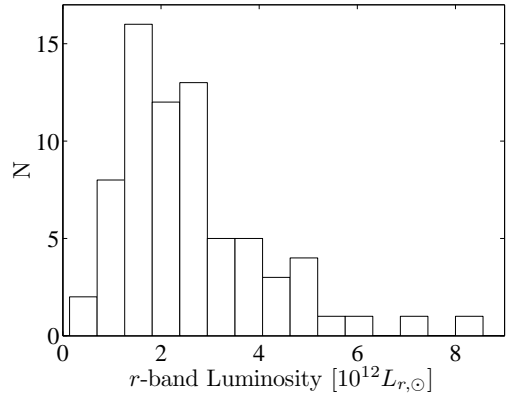}
\caption{Distribution of stellar luminosities of the 
WOOTS clusters that are in the SDSS DR4 (the ``SDSS clusters''),
as measured with our photometric aperture technique. \label{fig.Lhist}}
\end{figure}

Table \ref{table.clusters} lists our luminosity measurements for the 72 clusters
in our sample that are covered by the SDSS DR4 (the ``SDSS clusters'').  
Due to the WOOTS selection criteria (Paper I), the luminosities span a small
range, with  $68\%$ of the clusters having $L_r=2.3^{+1.7}_{-0.9} \times
10^{12}L_{r,\sun}$ (Fig.~\ref{fig.Lhist}). Within this range, there is no clear
trend with other cluster parameters. We therefore assign to the clusters for
which we do not have SDSS data (the ``non-SDSS clusters'') random luminosities
drawn from the luminosity distribution of the SDSS clusters. 
We estimate the uncertainties in the SN rate due to this random luminosity
assignment in \S~\ref{S.results}. 

The actual luminosity that we require in Eq.~\ref{eq.SNR} is not the total
cluster luminosity, but the cluster luminosity included within the search area.
For every WOOTS image we define the effective search area as the overlap
area between the image and its reference image. The reference image was assigned
a zero search area, since the survey was sensitive only to SNe that are brighter
in the later epoch (see Paper I). 
For the SDSS clusters, we calculate the luminosity, as described
above, contributed by 
galaxies that are inside the effective area of an image pair. 
If the full cluster area within
$R_c$ is covered by the image (usually clusters with $R_c<6'$), we adopt the
calculated cluster luminosity instead.  

\begin{figure}[ht]
\plotone {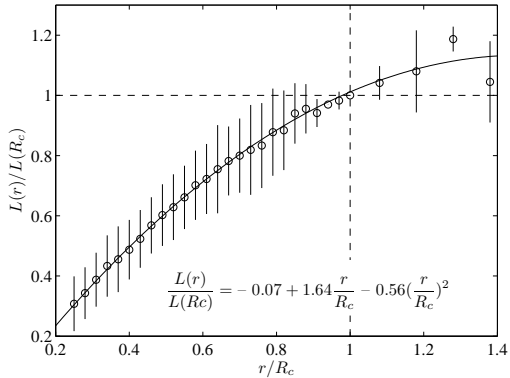}
\caption{Composite enclosed cluster-galaxy light profile, based on all SDSS
clusters in the sample. Error bars represent the measured rms over the sample.  
The solid line is the indicated second-order polynomial fit to the data. \label{fig_ia}}
\end{figure}

\begin{figure}[ht]
\plotone {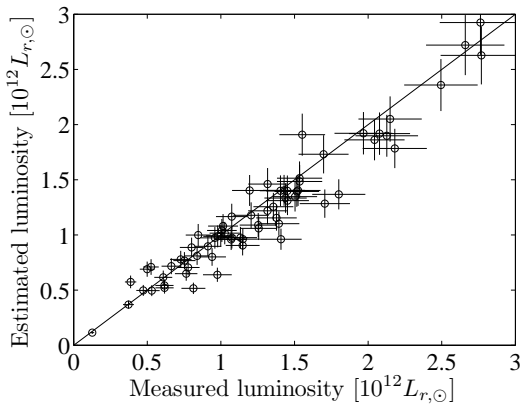} 
\caption{Comparison between the luminosity within the WOOTS search area,
 estimated using the relation in Fig.~\ref{fig_ia}, and the actual luminosity
measured in the image area, for the SDSS clusters.  
The solid line represents perfect agreement. \label{fig_Limg}}
\end{figure}

To estimate the luminosity inside the search area of the non-SDSS clusters, we
need to scale down the cluster luminosity according to the luminosity profile
and the effective search area. 
Figure \ref{fig_ia} shows, for the SDSS clusters, the mean fraction of cluster
luminosity, $L(<r)/L(<R_c)$, measured within an aperture of radius $r$, vs. the
radius normalized by the cluster radius, $r/R_c$. The solid line is a
second-order polynomial fit to the data, $L_r/L_{R_c} = -0.07 + 1.64r/R_c
-0.56(r/Rc)^2$. We use this relation for all the non-SDSS clusters with
$R_c>6'$, adopting for $r$ the radius of a circle with an area equal to the
effective search area of the image. 
Figure \ref{fig_Limg} examines the reliability of this approximation. For each
SDSS cluster, we estimate the luminosity within a circular aperture with an area
of $12'\times12'$ (the dimensions of the CCD used in WOOTS), using the above
relation, and plot it against the luminosity that is measured directly in the
effective search area. The difference between the estimated luminosities and the
measured ones has an rms deviation of $16\%$, which we adopt as the uncertainty
introduced by this approximation. 

\section{Results and Error Estimation}\label{S.results}

Combining all the elements of the calculation, as described above, we can derive
the cluster SN rate and its uncertainty. 
The dominant source of error in the rate is the Poisson statistics of the
(small) number of SNe, $N=6^{+3.58}_{-2.38}$ (68\% confidence limit). 
To determine the propagation of the uncertainties in all the parameters and
measurements that enter the rate derivation to the overall systematic
uncertainty, we have conducted a Monte Carlo experiment, in which we calculate
the SN rate many times, with the parameters drawn at random each time from their
respective distributions. The distributions are either Gaussian with an rms
deviation set to equal the parameter error, or a specific measured distribution
(e.g., the distribution of light-curve stretch factors).  
The distributions of SN rates from the Monte Carlo simulation permit both a
reliable error estimation and a determination of the most probable rate. Because
of the nonlinear dependence of the rate on some of the parameters entering the
calculation (e.g., stretch factor and luminosity), the most probable SN rate
will not necessarily be the SN rate obtained from the combined most-probable
values of all the input parameters. 
We will take as the most probable SN rate the peak of the Monte Carlo
distribution, with a systematic error derived directly from this distribution,
and a statistical error that originates from the Poisson confidence limits on
the number of events. 
 
We have also examined the sensitivity of the SN rate to the uncertainties in the
individual parameters, by turning on the Monte Carlo simulation for each one
separately.  
Table~\ref{table.std} summarizes the mean values and the uncertainty
in the SN rate, due to each of the parameters. \\
{\bf Luminosity error:} The overall luminosity error is about $5\%$, and is
dominated by the process of randomly drawing the non-SDSS cluster luminosities
from the luminosity distribution of the SDSS clusters. Varying the cluster
radius has a much smaller effect, of $\sim 1\%$.
Similarly, the uncertainty due to the luminosity gradient of the clusters
introduces an SN rate error smaller than $0.2\%$. Varying the inner radius of the
background annulus results in an error of less than $0.5\%$. \\ 
{\bf Visibility time error:} The  uncertainties in the visibility time
measurement have a less intuitive effect on the final rate distribution. The
visibility time is most sensitive to the shape and parameters of the efficiency
function, mainly $m_{0.5}$, which is the magnitude at which the efficiency drops
to half its maximum value.  
We find that drawing the efficiency parameters from normal distributions
centered on their best estimated or mean values results not only in a dispersion
in SN rates, but also in an overall decrease in the rate. The reason for this
behavior is that the dependence of the rate on the efficiency is nonlinear. A
slightly higher efficiency (caused by higher $m_{0.5}$ or lower $s$) will result
in a decrease in the SN rate, and vice versa, but for a similar absolute change
in the efficiency parameters, the decrease in the rate is larger than the
increase in the rate. The same applies to the distribution of SN light-curve
stretch factors. Drawing light curves from a stretch-factor distribution, even
if it is symmetric about the mean value, causes an overall decrease in the mean
SN rate.  
In our case, all of these effects are, of course, negligible compared to the
statistical error.  
The total systematic error, which in our case is $\sim6\%$, will become
comparable to the statistical error in future surveys with the detection of
several hundred SNe. 

\begin{table*}[ht]
\begin{center}
\caption{Sensitivity of the SN Rate to the Uncertainty in Individual Parameters
and Measurements.\label{table.std}} 
\begin{tabular}{llll}
\tableline\tableline 
Parameter                        & Parameter distribution     &	 $<SNu_r>$   & $\sigma$ \\ 
\tableline 
Non-SDSS cluster-luminosity draw & SDSS cluster               &  $0.325$   & $0.016$ \\ 
                                 & luminosity distribution    &            &         \\ 
Cluster radius ($R_c$)           & Normal with $\sigma=20\%$  &  $0.327$   & $0.004$ \\ 
$L$(in image) $\propto L(<r)$    & Normal with $\sigma=0.16$  &  $0.324$   & $0.002$ \\ 
Background annulus               & Measured$^1$; $\sigma=0.07$&  $0.3243$   & $0.0007$ \\ 
\tableline
Overall uncertainty in $L$       &                            &  $0.327$   & $0.017$ \\ 
\tableline\tableline
ZP (from USNO-B / SDSS)   & Normal with $\sigma=0.3 / 0.04$   &  $0.316$   & $0.004$ \\ 
$m_{0.5} vs. ZP$                 & Normal with $\sigma=0.4$   &  $0.301$   & $0.007$ \\ 
$s vs.  N_{Res}$                 & Normal with $\sigma=0.3$   &  $0.321$   & $0.002$ \\ 
LC stretch factor          & From Sullivan et al. (2006b) &  $0.318$   & $0.004$ \\ 
\tableline
Overall uncertainty in visibility time
                                 &                            &  $0.287$   & $0.009$ \\ 
\tableline\tableline	
Overall uncertainty from all parameters     &                 &  $0.290 $  & $0.017$ \\ 
\tableline\tableline		\\
\end{tabular}

\tablenotemark{1}{By calculating the luminosity for different buffer selections; see \S~\ref{s.luminosity}}
\end{center}
\end{table*}

We note that some parameters were not varied in our simulation, since their
errors are significantly smaller.  
(1) The uncertainties in the Schechter luminosity function parameters cause a
symmetrical change of $0.16\%$ in the luminosity, in the most extreme case, and
were therefore ignored. (2) In the calculation of cluster luminosities, we
assumed that all of the cluster light is emitted by elliptical galaxies.
However, a small fraction of the light does come from spiral galaxies, which
have a different K-correction -- it is $0.16$~mag smaller at $z=0.06$,
and $0.26$~mag larger at $z=0.2$. To check the effect of this assumption,
we calculated the SN rate for the unrealistic case, in which half of the cluster
galaxies are Sc galaxies. In the $0.06< z< 0.2$ range, individual cluster
luminosities increase (at $z<0.115$) or decrease (at $z>0.115$) by no more than
$10\%$, relative to the all-elliptical case. The actual distribution of cluster
redshifts leads to  a $\sim4\%$ increase in the SN rate. Thus, even in this
extreme scenario, the influence on the SN rate is smaller than that of the other
sources of systematic error.  

The resulting SN rates, as determined from the positions of the peaks of the
Monte Carlo distributions, and normalized by stellar luminosity in the $g,r,i$,
and $z$ bands are given in Table~\ref{table.SNR}. The rates are given in units
of SNu$_{\rm band}$, defined as  SNe (century $10^{10} L_{{\rm
band},\sun}$)$^{-1}$. We also express our result in the Johnson $B$ band used
traditionally for SN rates, in units of SNu$_B$, by assuming $B-r=1.078$ mag for
elliptical galaxies (as calculated from the template elliptical spectrum of
Kinney et al. 1996), $M_{r,\sun}=4.64$ (Blanton et al., 2006), and
$M_{B,\sun}=5.47$ (Allen 1976). 

\begin{table*}[ht]
\begin{center}
\caption{SN~Ia Rates\label{table.SNR}}
\begin{tabular}{lllll}
\tableline\tableline
Environment  & Redshift      & Value $^1$                             & Units$^2$   & Reference \\ 
\tableline 
cluster      & $0.06<z<0.2$  & $0.351 ^{+0.210}_{-0.139} \pm 0.020 $  & SNu$_g$     & This work \\ 
             &               & $0.288 ^{+0.172}_{-0.114} \pm 0.018 $  & SNu$_r$     &           \\ 
             &               & $0.229 ^{+0.137}_{-0.091} \pm 0.014 $  & SNu$_i$     &           \\ 
             &               & $0.186 ^{+0.111}_{-0.074} \pm 0.010 $  & SNu$_z$     &           \\ 
\tableline 
cluster      & $0.06<z<0.2$  & $0.36 ^{+0.22}_{-0.14} \pm 0.02 $      & SNu$_B$     & This work \\ 
cluster      & $0.18<z<0.37$ & $0.39 ^{+0.59}_{-0.25}$                & SNu$_B$     & Gal-Yam et al. (2002) \\ 
cluster      & $0.83<z<1.27$ & $0.80 ^{+0.92}_{-0.41}$                & SNu$_B$     & Gal-Yam et al. (2002) \\ 
E/S0         & local         & $0.16 \pm 0.05        $                & SNu$_B$     & Capellaro et al (1999) \\ 
\tableline 
cluster      & $0.06<z<0.2$  & $0.098 ^{+0.059}_{-0.039} \pm 0.009$   & SNuM        & This work \\ 
E/SO         & local         & $0.038 ^{+0.014}_{-0.012}$             & SNuM        & Mannucci et al. (2005)\\ 
E/SO         & $0.2<z<0.75$  & $0.053\pm{0.011}$                      & SNuM        & Sullivan et al. (2006)\\ 
\tableline 
\end{tabular}\\
\tablenotetext{1}{Where two sets of errors are presented, they are statistical and
systematic, respectively.}
\tablenotetext{2}{SNu$_{\rm band}$ $\equiv$ SNe (100 yr)$^{-1}$ ($10^{10} {\rm L}_{{\rm
band},\sun})^{-1}$.\\ 
~SNuM $\equiv$ SNe (100 yr)$^{-1}$ ($10^{10} {\rm M}_{\sun})^{-1}$}
\end{center}
\end{table*}
 
\section{Comparison to Other SN Rate Measurements}\label{S.comparison}
Our derived cluster SN rate at $0.06<z<0.19$ can be compared to recent
measurements of SN rates both in clusters and in the field.  
In the field, we compare the rate derived in this work to recent measurements of
the SN rate in elliptical galaxies per unit stellar luminosity and per unit
mass. 
Cappellaro et al. (1999) measured a local E/S0 SN~Ia rate of $0.16\pm
0.05$~SNu$_{B}$, which is consistent with the  rate we obtain in this luminosity
band, \snuB~SNu$_B$.   
To convert our measurement to SNuM [SNuM = SN (century $10^{10}
M_{\sun}$)$^{-1}$], we follow Mannucci et al. (2005), who used the mass-to-light
ratio derived by Bell et al. (2001) to convert the $K$-band luminosities and
$B-K$ colors of the galaxies in their sample to stellar mass. 
We derive the $r-i$ color of each SDSS cluster from the ratio between its
measured $r$-band and $i$-band luminosities, $r-i=-2.5\,{\rm
log}_{10}(L_r/L_i)-M_{i,\sun}+M_{r,\sun}$. The stellar mass of each cluster is
then estimated from the color-dependent stellar mass-to-light ratio derived by
Bell et al. (2003), ${\rm log}_{10}(M_{\sun}/L_{z,\sun})= -0.052 + 0.923(r-i)$
(see Manucci et al. 2005 for a discussion of the validity of this ratio for our
purpose).  
Finally, we define the stellar mass-to-light ratio of our cluster sample as the
total mass divided by the total $z$-band luminosity of the entire sample, $M/L =
1.89 \pm 0.44 M_\sun/L_{z,\sun}$, and use it to convert the SN rate from SNu$_z$
to SNuM. The resulting SN rate per unit stellar mass is \snuM~ SNuM. 
Mannucci et al. (2005) used the SN sample of Cappellaro et al. (1999) to measure
the SN rate per unit mass as a function of host-galaxy morphological type. They
found a rate of $0.038 ^{+0.014}_{-0.012}$ (converted from $h=0.75$ to $h=0.70$)
SNuM for low-redshift ($z<0.02$) E/S0 galaxies, lower than, but consistent with our result for 
cluster galaxies at a somewhat higher redshift.  
We note that some of the E/S0 galaxies monitored by Cappellaro et al. are members 
of nearby galaxy clusters, and therefore the Mannucci et al. (2005) local rate
includes both cluster and field early-type galaxies. A discussion of local rates 
separated by environment will be presented elsewhere.

Sullivan et al. (2006) measured the SN~Ia rate at $0.2<z<0.75$, from $124$ SNe
discovered by the Supernova Legacy Survey (SNLS), and studied the SN host
properties. Following Mannucci et al. (2005) and Scannapieco et al. (2006), they
separated the rate into two components, one proportional to stellar mass, and
the other proportional to the star formation rate --
$SNR_{Ia}(t)=AM(t)+B\dot{M}(t)$. Their best-fit values are $A=0.053\pm{0.011}$
SNuM, $B=(3.9 \pm 0.7)\times 10^{-4}$ SNe yr$^{-1}$ (M$_\sun$ yr$^{-1}$)$^{-1}$.
Since the stellar mass in their sample is dominated by the early-type galaxies,
and these galaxies have virtually no star formation, the parameter $A$ is
essentially the SN rate in E/S0 galaxies. Comparison of the E/S0 SN rates
measured by Mannucci et al. (2005) and Sullivan et al. (2006) indicates that the
rate may be constant  with redshift out to $z \approx 0.5$, although an increase by
a factor $\sim 2$ is also consistent within the errors. Our measurement in
clusters at an intermediate redshift range is consistent with these two field
elliptical rate measurements.

In galaxy clusters, Gal-Yam et al. (2002) measured cluster SN~Ia rates per unit
stellar $B$-band luminosity of $0.39^{+0.59}_{-0.25}$ SNu$_B$ at  $<z>=0.25$ and
$0.80^{+0.92}_{-0.41}$ SNu$_B$ at $<z>=0.9$ (converting to our  adopted
cosmology). These estimates were based on the detection of one cluster SN in the
low-redshift bin, and one or two in the high-redshift bin, and consequently, the
errors are large. 
Comparing to our measurement of  \snuB~SNu$_B$, there is a hint for a slow rise
in SN rate with redshift. However, the data are equally consistent, given the
large error bars, with a constant rate. More accurate SN rate measurements at
high $z$, which are in progress, will elucidate this question.

\section{Summary}
We have analyzed the data from the WOOTS cluster SN survey (Paper I) to derive
the SN~Ia rate in $0.06<z<0.19$ clusters, normalized by stellar luminosity
and by stellar mass (Table~\ref{table.SNR}).  
In addition to the sample of six cluster SNe~Ia discovered by the survey, our
measurement required the determination of two variables -- the survey visibility
time, and the cluster stellar luminosities, which we determined using the survey
data, assisted by SDSS data.   
We have conducted Monte Carlo simulations that mimic the inhomogeneity in the
parameters that enter the rate calculation, and quantify the dependence of the
systematic errors on the uncertainty in these parameters. The resulting
systematic errors are about an order of magnitude smaller than the Poisson
errors due to the small number of SNe that were detected. 

We find that the SN rate is similar to the rate found in field elliptical
galaxies, both locally and out to $z=0.75$. A comparison to cluster SN rates at
higher redshifts is similarly consistent with an unchanging rate, but the large
uncertainties in current high-$z$ cluster SN rates can also accommodate a 
decline in the rate with time.  
 We are in the process of obtaining cluster SN rate measurements at high redshift.
The emerging dependence of SN rate on cosmic time and on galaxy environment
should provide valuable new insights for astrophysics and cosmology. 


\acknowledgments
We thank Derek Fox for assistance with the image calibration, Dovi Poznanski for help
with synthetic photometry, and Eran Ofek and Ben Koester for numerous useful
discussions.  
We are grateful to Louis DesRoches, 
Chung-Pei Ma, Michael Blanton, and David Hogg, for input
regarding BCG luminosities in the SDSS. The anonymous
referee is thanked for useful comments and suggestions.
This work was supported by a grant from the Israel Science Foundation (D.M.). 
A.G. acknowledges support
by NASA through Hubble Fellowship grant \#HST-HF-01158.01-A awarded by
STScI, which is operated by AURA, Inc., for NASA, under contract NAS
5-26555. A.V.F. is grateful for the support of National Science Foundation
grants AST-0308 and AST-0307894 and AST-0607485.



\end{document}